\pdfoutput=1
\documentclass{emulateapj}
\usepackage{rotating}
\usepackage{graphicx}

\slugcomment{}

\shorttitle{Triggered star formation around mid-infrared bubbles in G8.14+0.23 H\,{\sc ii} region}
\shortauthors{L.~K. Dewangan et al.}

\begin{document}


\title{Triggered star formation around mid-infrared bubbles in G8.14+0.23 H\,{\sc ii} region}
\author{L.~K. Dewangan\altaffilmark{1}, D.~K. Ojha\altaffilmark{1}, B.~G. Anandarao\altaffilmark{2}, S.~K. Ghosh\altaffilmark{3}, and S. Chakraborti\altaffilmark{1} }

\email{lokeshd@tifr.res.in}

\altaffiltext{1}{Department of Astronomy and Astrophysics, Tata Institute of Fundamental Research, Homi Bhabha Road, Mumbai 400 005, India}
\altaffiltext{2}{Physical Research Laboratory, Navrangpura, Ahmedabad - 380 009, India}
\altaffiltext{3}{National Centre for Radio Astrophysics, Ganeshkhind, Pune - 411 007, India}




\begin{abstract}
Mid-infrared (MIR) shells or bubbles around expanding H\,{\sc ii} regions have received much attention due to their ability to 
initiate a new generation of star formation. We present multi-wavelength observations around two bubbles associated with 
a southern massive star-forming (MSF) region G8.14+0.23, to investigate the triggered star formation signature on the 
edges of the bubbles by the expansion of the H\,{\sc ii} region. We have found observational signatures of the collected molecular and 
cold dust material along the bubbles and the $^{12}$CO(J=3-2) velocity map reveals that the molecular gas in the bubbles is physically 
associated around the G8.14+0.23 region. 
We have detected 244 young stellar objects (YSOs) in the region and about 37\% of these YSOs occur in clusters. 
Interestingly, these YSO clusters are associated with the collected material on the edges of the bubbles. 
We have found good agreement between the dynamical age of the H\,{\sc ii} region and 
the kinematical time scale of bubbles (from the $^{12}$CO(J=3-2) line data) with the fragmentation 
time of the accumulated molecular materials to explain possible ``collect-and-collapse" process around the G8.14+0.23 region. 
However, one can not entirely rule out the possibility of triggered star formation by compression of the pre-existing dense 
clumps by the shock wave. We have also found two massive embedded YSOs (about 10 and 22 M$_{\odot}$) which are associated with 
the dense fragmented clump at the interface of the bubbles. We conclude that the expansion of the H\,{\sc ii} region is also 
leading to the formation of these two young massive embedded YSOs in the G8.14+0.23 region. 
\end{abstract}


\section{Introduction}
\label{sec:intro}
Massive stars have a major impact on their surrounding environment due to their energetic wind, UV ionizing radiation and 
an expanding H\,{\sc ii} region \citep{zinnecker07}. 
In recent years, MIR shells or bubbles around the expanding H\,{\sc ii} regions have
received much attention, because of their ability to initiate a new generation of star formation \citep{watson08,watson09,watson10}.
\citet{churchwell06,churchwell07} identified about 600 such shells or bubbles based on a 
circular morphology in {\it Spitzer}-GLIMPSE 8 $\mu$m emission. It is now believed that such regions are very promising 
to observationally study the conditions of sequential/triggered star formation \citep[and references therein]{thompson12}.

In this paper, we present a study of two MIR bubbles, CN107 and CN109, from the catalogue of \citet{churchwell07} around the G8.14+0.23 region. 
The G8.14+0.23 is an irregular Galactic ultra-compact (UC) H\,{\sc ii} region close to the IRAS 17599-2148 source, situated at a 
distance of 4.2 kpc \citep{kim01}. The G8.14+0.23 region has been extensively studied in the radio, sub-mm and molecular line 
observations \citep{codella94,walsh97,walsh98,kim01,kim03,ojeda02,thompson06}. 
The region is ionized by an O6(B0) spectral class source and is also known as a bipolar blister type H\,{\sc ii} region due to a champagne flow \citep{kim01,kim03}. 
The 6.7 GHz Class II methanol maser \citep{walsh97,walsh98} and  water maser \citep{codella94} were also 
detected close to the IRAS position. The site of massive star formation is often associated with methanol and water maser emission.
\citet{kim01} found the velocity of ionized gas to be about 20.3 km s$^{-1}$ close to the IRAS position, using H76$\alpha$ recombination 
line profile. The velocity of molecular flow was obtained to be about 19.6 km s$^{-1}$ using CS J= 2-1 lines by \citet{kim03}. 
These velocities are close to the values obtained by NH$_{3}$(2,2) and (4,4) emissions to be about 19.3 km s$^{-1}$ \citep{churchwell90} 
and 20.3 km s$^{-1}$ \citep{cesaroni92}, respectively. \citet{kim03} found irregular morphology and clumpy structure of molecular cloud 
around G8.14+0.23, using $^{13}$CO (J= 1-0) and CS (J= 2-1) line intensity maps. \citet{kim03} and \citet{thompson06} also found a molecular 
and dust emission ridge around the region. \citet{okada08} analyzed the MIR spectroscopic observations using $\it Spitzer$-Infrared 
Spectrograph (IRS) for 14 Galactic star-forming regions, including G8.14+0.23, in the wavelength range of 20 - 36.5 $\mu$m. 
They argued the presence of the shocked and ionized gas associated with the G8.14+0.23 region, based on the detection of [Si\,{\sc ii}], [Fe\,{\sc ii}] 
and [S\,{\sc iii}] emission lines and derived the abundance ratio of two ions against the solar abundance ({\it as}) as 
0.060, 0.158 and 0.020 for (Fe$^{+}$/Si$^{+}$)$_{as}$, (Si$^{+}$/S$^{2+}$)$_{as}$ and (Fe$^{2+}$/S$^{2+}$)$_{as}$ respectively.
 
In recent years, multi-wavelength observations are utilized to trace out different components associated with the 
star-forming regions viz. ionized, shocked, cold dust and molecular gas, as well as embedded young stellar populations. 
In this paper, we present multi-wavelength observations to investigate the triggered star formation around the MIR bubbles 
associated with a southern MSF region G8.14+0.23. This region is not well explored in infrared wavelengths, especially the 
identification of embedded populations around the region. 
Our study also includes the search for the infrared counterpart(s) of embedded massive YSOs in their earlier phases and to understand their 
formation processes. 

In Section~\ref{sec:obser}, we introduce the archival data and data reduction procedures used for the present study. 
In Section~\ref{sec:data}, we examine the morphology of the G8.14+0.23 region in different wavelengths and 
the interaction of massive stars with its environment using various {\it Spitzer} MIR ratio maps. 
We also describe the selection of young population, their distribution around the G8.14+0.23 region and 
discuss the triggered star formation scenario on the edge of the bubbles in this section. 
In Section~\ref{sec:conc}, we summarize our conclusions. 
\section{Available data around G8.14+0.23 and data reduction}
\label{sec:obser}
Archival deep  near-infrared (NIR) JHK$_{s}$ images and catalogue of G8.14+0.23 region were obtained from the 
UKIDSS 6$^{th}$ archival data release (UKIDSSDR6plus) of the Galactic Plane Survey (GPS) \citep{lawrence07}.
UKIDSS observations were made using the UKIRT Wide Field Camera \citep[WFCAM;][]{casali07} 
and fluxes were calibrated using Two Micron All Sky Survey \citep[2MASS;][]{skrutskie06}. 
The details of basic data reduction and calibration procedures are described in \citet{dye06} and \citet{hodgkin09}, respectively. 
Magnitudes of bright stars (J $\leqslant$ 11.5 mag, H $\leqslant$ 11.5 mag and K$_{s}$ $\leqslant$ 10.5 mag) were obtained from the 2MASS, 
due to saturation of UKIDSS bright sources. 
Only those sources are selected for the study which have photometric magnitude error of 0.1 and less in each band.
 
We obtained narrow band molecular hydrogen (H$_{2}$; 2.12 $\mu$m; 1 - 0 S(1)) imaging data from UWISH2 survey \citep{froebrich11}. 
We followed a similar procedure as described by \citet{varricatt11} to obtain the final continuum-subtracted H$_{2}$ image 
using GPS K$_{s}$ image.
 
The {\it Spitzer} Space Telescope Infrared Array Camera (IRAC (Ch1 (3.6 $\mu$m), Ch2 (4.5 $\mu$m), Ch3 (5.8 $\mu$m) and Ch4 (8.0 $\mu$m); \citet{Fazio04}) and 
Multiband Imaging Photometer (MIPS (24 $\mu$m); \citet{rieke04}) archival images were obtained around the G8.14+0.23 region 
from the ``Galactic Legacy Infrared Mid-Plane Survey Extraordinaire'' (GLIMPSE) and ``A 24 and 70 Micron Survey of the Inner Galactic Disk with MIPS'' 
(MIPSGAL) surveys. MIPSGAL 24 $\mu$m image is saturated close to the IRAS 17599-2148 position. 
We performed aperture photometry on all the GLIMPSE images (plate scale of 0.6 arcsec/pixel) using a 2.4 arcsec aperture and a 
sky annulus from 2.4 to 7.3 arcsec using IRAF\footnote[1]{IRAF is distributed by the National Optical Astronomy Observatory, USA}. 
The photometry is calibrated using zero magnitudes including aperture corrections, 18.5931 (Ch1), 18.0895 (Ch2), 17.4899 (Ch3) and 16.6997 (Ch4), obtained 
from IRAC Instrument Handbook (Version 1.0, February 2010). 

We obtained 20 cm radio continuum map (resolution $\sim$ 6 arcsec) from Very Large Array (VLA) Multi-Array Galactic Plane Imaging Survey (MAGPIS) 
\citep{helfand06} to trace the ionized region around G8.14+0.23. 
The molecular $^{12}$CO(J=3-2) (rest frequency 345.7959899 GHz) spectral line public processed archival data was also utilized 
in the present work. The CO observations were taken on 11 June 2009 at the 15 m James Clerk Maxwell Telescope (JCMT) using the HARP array. 
The JCMT archival reduced fits cube was used to extract the velocity information around the region using the Astronomical Image Processing 
System\footnote[2]{http://www.aips.nrao.edu/} (AIPS) package. 
The fits cube was loaded into AIPS and rotated to make the frequency axis the first one. 
An alternate velocity axis was defined from the frequency using the rest frame
frequency of the CO J=3-2 transition. The data were Hanning smoothed over 476 m s$^{-1}$ to decrease the noise per channel. 
It was then integrated over the velocity range 10 to 30 km s$^{-1}$ to obtain the zeroth moment (column density), 
first moment (mean velocity) and second moment (velocity dispersion). 
Archival BOLOCAM 1.1 mm \citep{aguirre11} and SCUBA 850 $\mu$m \citep{francesco08} images were also used in the present work. 
The effective FWHM Gaussian beam sizes of final 1.1 mm and 850 $\mu$m images were 33 arcsec and 22.9 arcsec, respectively. 
\section{Results and Discussion}
\label{sec:data}
\subsection{Morphology of the G8.14+0.23 region in different wavelengths}
\label{subsec:morpho}
\begin{figure}
\plotone{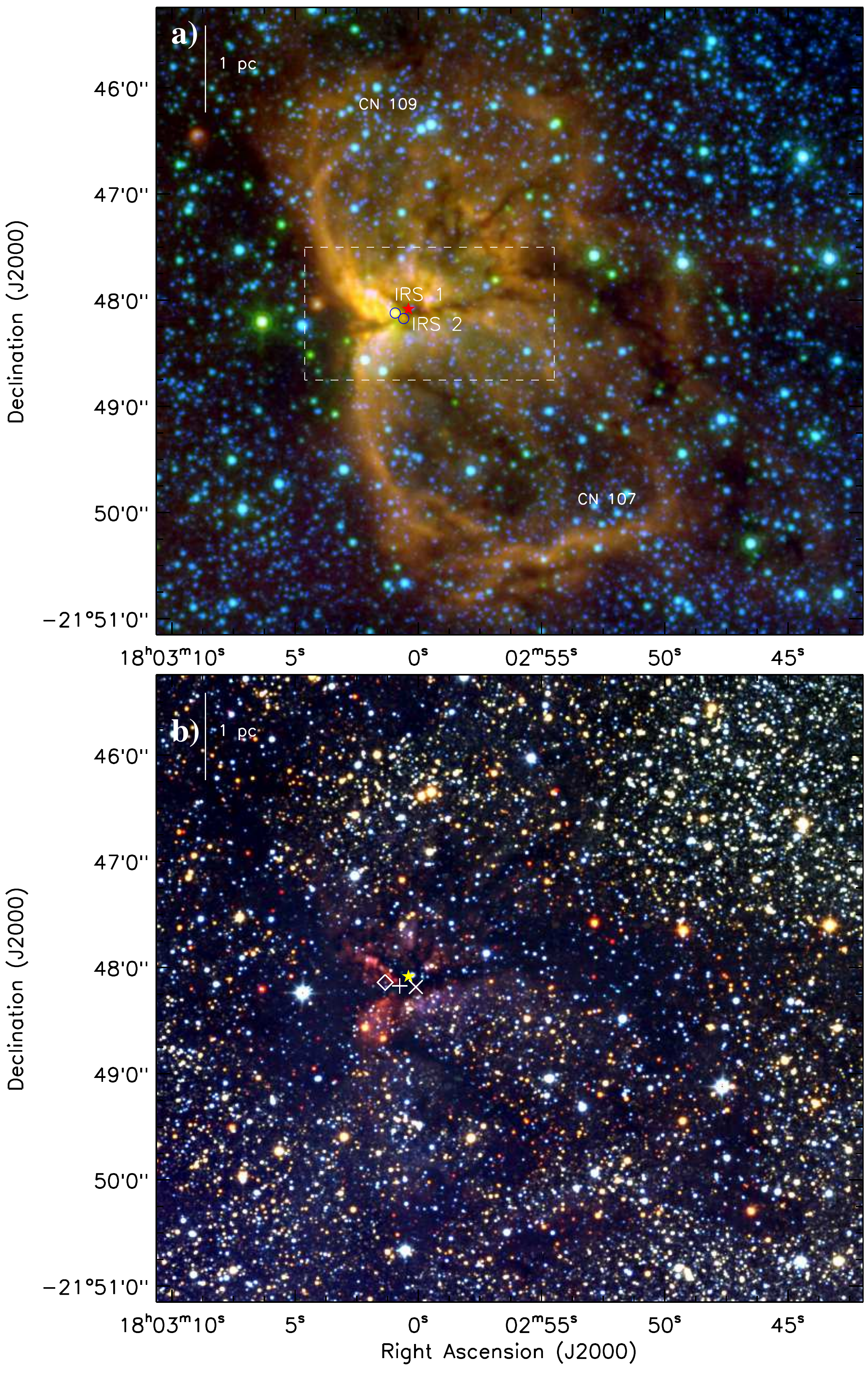}
\caption{\footnotesize\scriptsize a) 3 color composite image (size $\sim$ 6.6 $\times$ 5.9 arcmin$^{2}$;
central coordinates: $\alpha_{2000}$ = 18$^{h}$ 02$^{m}$ 56$^{s}$.3, $\delta_{2000}$ = -21$^{\degr}$ 48$^{\arcmin}$ 12$^{\arcsec}$.6)
of the G8.14+0.23 region, using {\it Spitzer}-GLIMPSE images at 8.0 $\mu$m (red), 4.5 $\mu$m (green) and UKIDSS K$_{s}$ (blue) in log scale.
The positions of the two sources (IRS~1 \& IRS~2) are marked by blue circles and labeled on the image.
The scale bar on the top left shows a size of
1 pc at the distance of 4.2 kpc. The white dashed box is shown as zoomed-in view in Fig.~\ref{fig8}.
Two bubbles from the \citet{churchwell07} are also labeled on the image as CN 107 and CN 109.
b) Color composite image using UKIDSS J (blue), H (green) and K$_{s}$ (red) in log scale.
The positions of IRAS 17599-2148 ($\bigstar$), UC~H\,{\sc ii} region ($\Diamond$), methanol maser (+) and water maser ($\times$) are
marked in the figure. \label{fig1}}
\end{figure}

\begin{figure}
\plotone{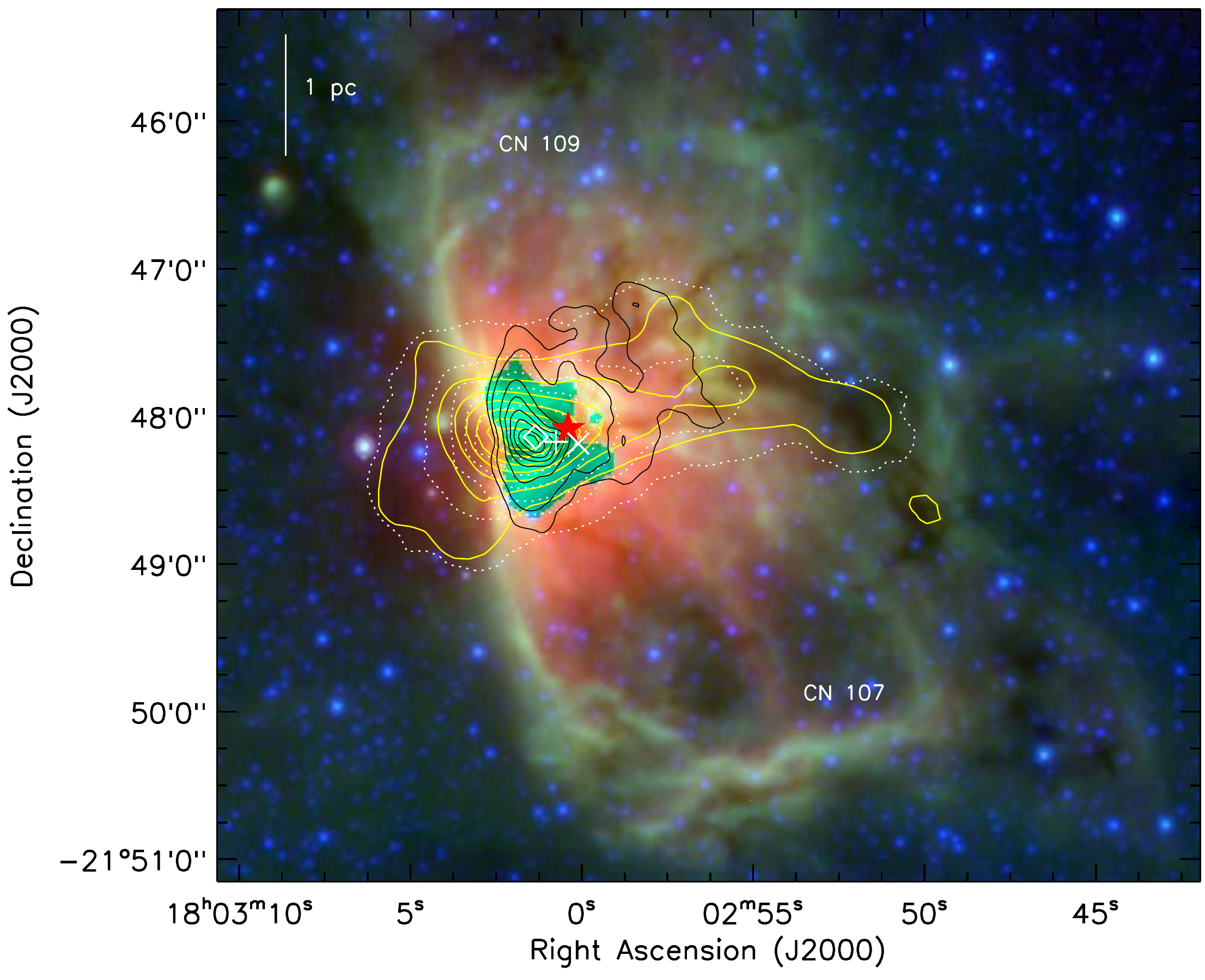}
\caption{Figure represents 20 cm contours in black color around the G8.14+0.23 region obtained from the MAGPIS survey,
overlaid on a color composite image made using the 24 $\mu$m (red), 8 $\mu$m (green), and 3.6 $\mu$m (blue) images.
The 20 cm contour levels are 10, 25, 40, 55, 70, 85 and 95 \% of the peak value i.e. 0.23 Jy/ beam.
BOLOCAM 1.1 mm and SCUBA 850 $\mu$m emissions are also shown by dotted white and solid yellow contours
with 10, 25, 40, 55, 70, 85 and 95 \% of the peak value i.e. 3.55 and 8.23 Jy/ beam, respectively.
Two bubbles from the \citet{churchwell07} are also labeled on the image as CN 107 and CN 109.
The marked symbols are similar to as shown in Fig.~\ref{fig1} (see text for more details).
MIPS 24 $\mu$m image is saturated near to the IRAS position. \label{fig2}}
\end{figure}

Figure~\ref{fig1}a represents the RGB color composite image using GLIMPSE (8.0 $\mu$m (red) \& 4.5 $\mu$m (green)) and UKIDSS K$_{s}$ (blue) 
of a region ($\sim$ 6.6 $\times$ 5.9 arcmin$^{2}$) around G8.14+0.23. The 8 $\mu$m band contains the two strongest polycyclic aromatic 
hydrocarbon (PAH) features at 7.7 $\mu$m and 8.6 $\mu$m, which are excited in
the photodissociation region  (or photon-dominated region, or PDR). The PDRs are the interface between neutral \& molecular hydrogen and traced 
by PAH emissions. Figure~\ref{fig1}b shows the 3 color composite image using UKIDSS J (blue), H (green) and K$_{s}$ (red) band 
in log scale. The positions of IRAS 17599-2148 ($\bigstar$), UC~H\,{\sc ii} region ($\Diamond$), methanol maser (+) 
and water maser ($\times$) are marked in the figure. The coordinates of UC~H\,{\sc ii} region, methanol maser and water maser 
were taken from \citet{kim01}, \citet{walsh98} and \citet{codella94}, respectively.
Figure~\ref{fig1}a displays the two extended MIR bubbles CN107 and CN109 prominently around the G8.14+0.23 region. 
It is to be noted that these bubble structures are not visible in any of the UKIDSS NIR images (see Figure~\ref{fig1}b), but dark regions are seen 
corresponding to the bubble structures in Figure~\ref{fig1}b. 
The two infrared sources (designated as IRS~1 and IRS~2) are seen close to the 
IRAS position and are also marked in Figure~\ref{fig1}a (see subsection~\ref{subsec:phot1} for more details). 
Figure~\ref{fig2} shows color composite image made using MIPSGAL 24 $\mu$m (red), GLIMPSE 8 $\mu$m (green), and 3.6 $\mu$m (blue) images, 
overlaid by BOLOCAM 1.1 mm and SCUBA 850 $\mu$m emission by dotted white and solid yellow contours, respectively. 
MIPS 24 $\mu$m image is saturated near to the IRAS position. 
MAGPIS 20 cm radio continuum emission is also overlaid in Figure~\ref{fig2} by black contours. 
The cold dust emission at 850 $\mu$m is very dense and prominent at the interface of the two bubbles. 
It is to be noted that the peaks of cold dust and ionized gas are concentrated near the IRAS 
position and interface of the bubbles. The 24 $\mu$m and 20 cm images trace the warm dust and ionized gas in the 
region, respectively. It is obvious from Figure~\ref{fig2} that the PDR region (traced by 8 $\mu$m) 
encloses 24 $\mu$m and 20 cm radio emissions inside the bubbles and indicates the presence of dust 
in and around the H\,{\sc ii} region \citep[see for example][for N10, N21, and N49 bubbles]{watson08}.

\begin{figure}
\plotone{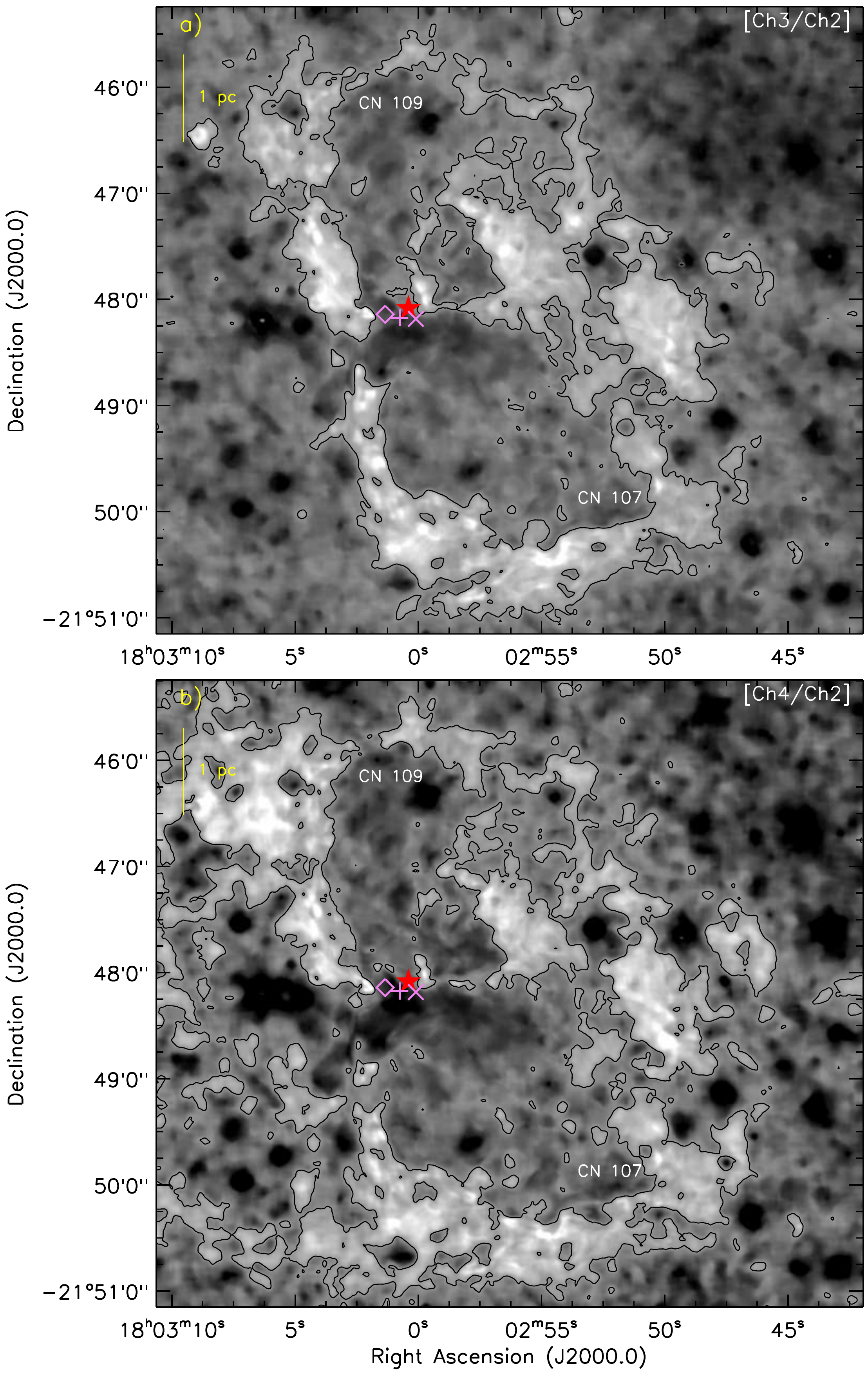}
\caption{\footnotesize\scriptsize a) Figure exhibits the GLIMPSE Ch3/Ch2 ratio map of the G8.14+0.23
region (similar area as shown in Fig.~\ref{fig1}).
The ratio Ch3/Ch2 value is found to be ``7--8'' and ``8.5--9'' in the bubble interior and for the brightest part of the PDR respectively.
The Ch3/Ch2 ratio contours are also overlaid on the image with a level of 7.7, a representative value between ``7--8''.
b) The Ch4/Ch2 ratio map of the region is shown here. The ratio Ch4/Ch2 value for the brightest part of the PDR and the interior
is ``30--34" and ``23--25" respectively. The Ch4/Ch2 ratio contours are also overlaid on the image with a level of 24.8, a
representative value between ``23--25''. The marked symbols are similar to as shown in Fig.~\ref{fig1}. \label{fig3}}
\end{figure}

\begin{figure}
\plotone{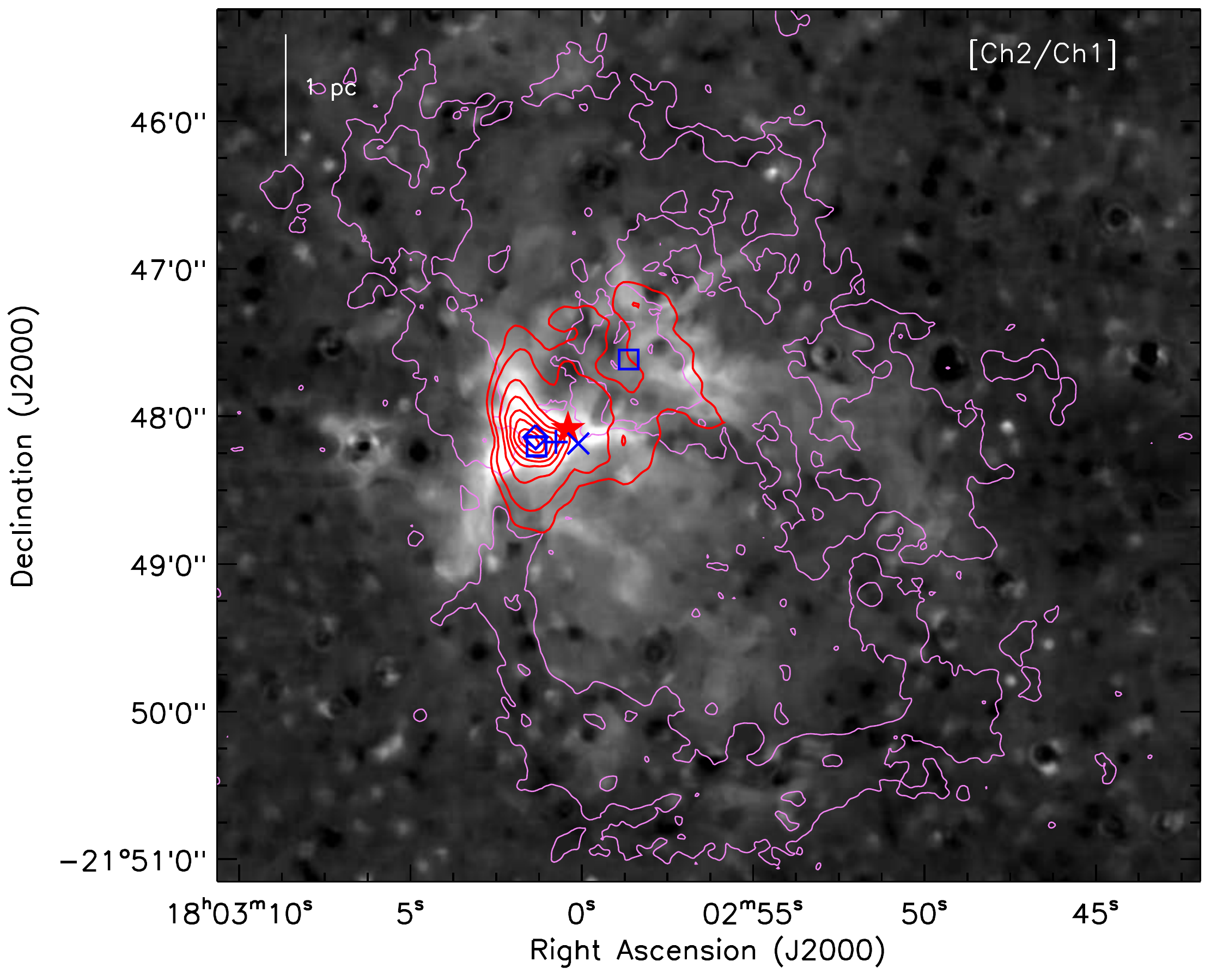}
\caption{Figure exhibits the Ch2/Ch1 ratio image of the G8.14+0.23 region.
The Ch3/Ch2 ratio contours in violet are also overlaid on the image with a level of 7.7.  Contours in red color represent
the 20 cm radio continuum emission with similar levels as shown in Fig.~\ref{fig2} around the region. Square symbols represent the 6 cm radio detections by \citet{ojeda02} and
other marked symbols are similar to as shown in Fig.~\ref{fig1}. \label{fig4}}
\end{figure}

\begin{figure}
\plotone{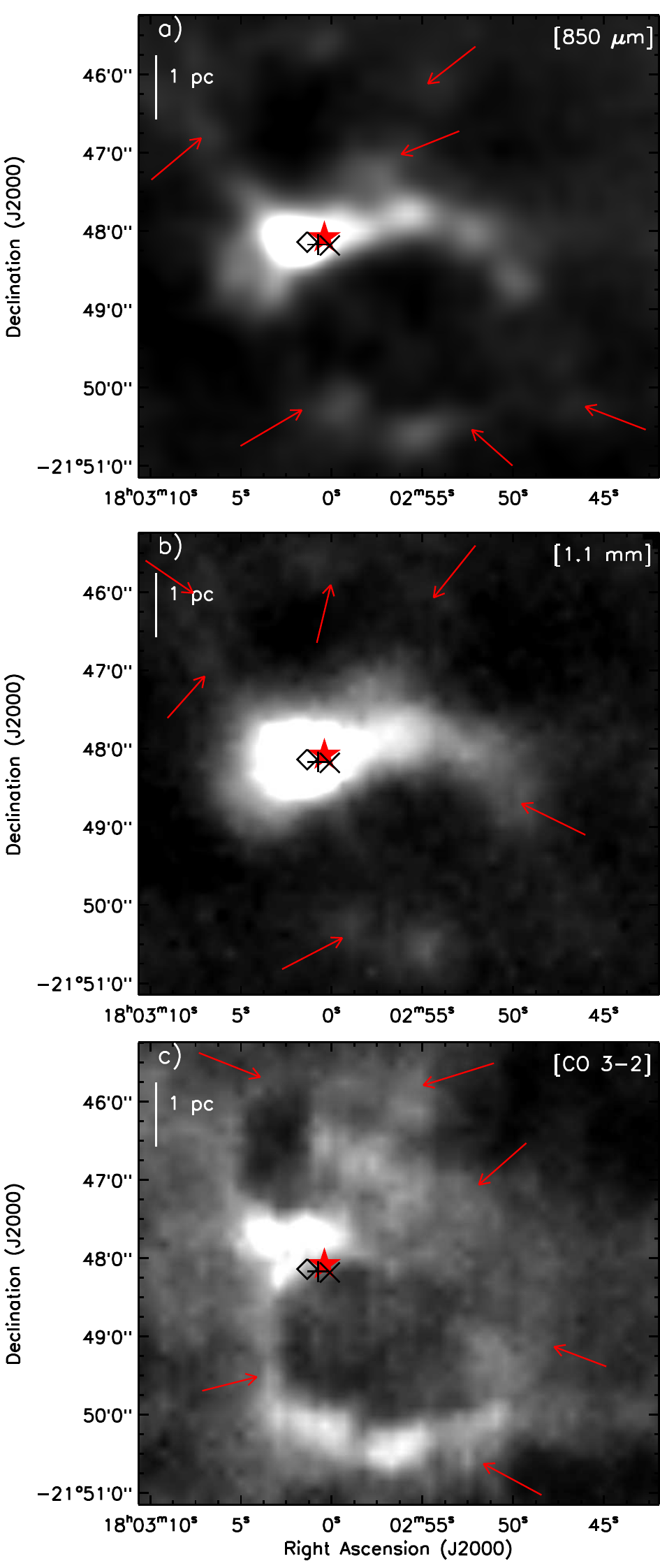}
\caption{The grey-scale image in SCUBA 850 $\mu$m (a), BOLOCAM 1.1 mm (b) and JCMT CO 3-2 (c) around the G8.14+0.23 region, respectively.
Red arrows show the detected faint features around the G8.14+0.23 region and other marked symbols are similar to as shown in Fig.~\ref{fig1}. \label{fig5}}
\end{figure}

\begin{figure}
\plotone{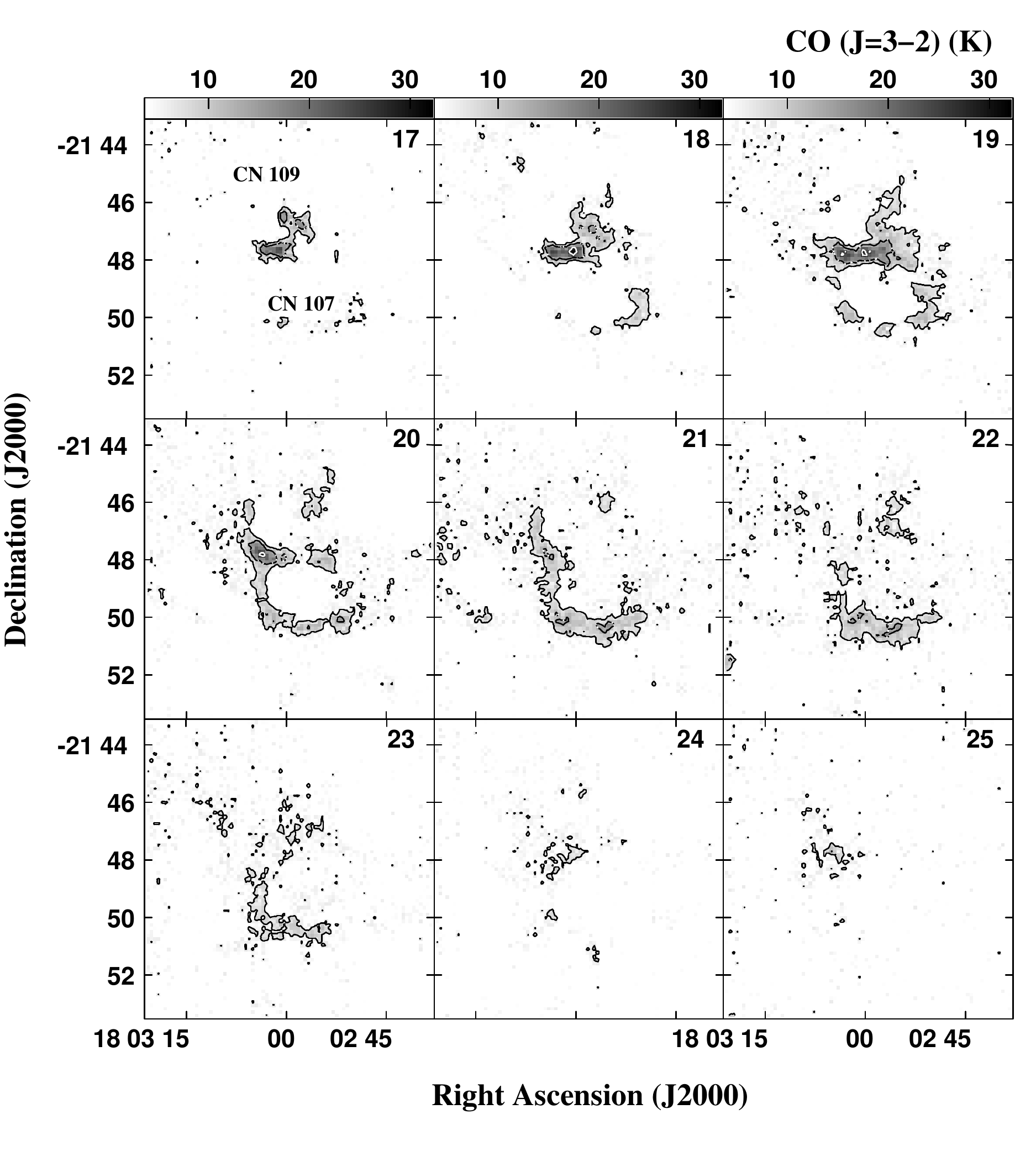}
\caption{Figure shows the $^{12}$CO(J=3--2) channel maps in 1 km s$^{-1}$ steps with velocity interval between 17 to 25 km s$^{-1}$
around the G8.14+0.23 region. The grey color scale bar at the top is presented in unit of K in the range of 4 to 32 K
and the contours are plotted for 6.4, 12.8 and 25.7 K, from outer to inner side. The bubbles CN 107 and CN 109 are also labeled in the top left panel. \label{fig6}}
\end{figure}

\subsection{IRAC Ratio Maps and the Collected material}
\label{subsec:ratmap}
In recent years, {\it Spitzer}-IRAC bands and ratio maps are utilized to study the interaction 
of massive stars with its immediate environment \citep{povich07}. The IRAC bands contain a number of prominent atomic and molecular lines/features such as 
H$_{2}$ lines in all channels \citep[see Table~1 from][]{smith05}, Br$\alpha$ 4.05 $\mu$m (Ch2), Fe\,{\sc ii} 5.34 $\mu$m (Ch3), Ar\,{\sc ii} 6.99 $\mu$m 
and Ar\,{\sc iii} 8.99 $\mu$m (Ch4) \citep[see][]{reach06}. It is to be noted that the {\it Spitzer}-IRAC bands, Ch1, Ch3 and Ch4, contain the PAH features 
at 3.3, 6.2, 7.7 and 8.6 $\mu$m, whereas Ch2 (4.5 $\mu$m) does not include any PAH features. 
Therefore, IRAC ratio (Ch4/Ch2, Ch3/Ch2 and Ch1/Ch2) maps are being used to trace out the PAH features in MSF 
regions \citep[e.g.][]{povich07,watson08,kumarld10} due to UV radiation from massive star(s). 
In order to make the ratio maps, we generate residual frames for each band removing point sources by 
choosing an extended aperture (12.2 arcsec) and a larger sky annulus \citep[14.6 - 24.4 arcsec;][]{reach05} in 
IRAF/DAOPHOT software \citep{stetson87}. These residual frames are then subjected to median filtering with a width of 10 pixels and smoothing 
by 3 $\times$ 3 pixels using the boxcar algorithm \citep[e.g.][]{povich07}. Figures~\ref{fig3}a and~\ref{fig3}b represent the IRAC ratio maps, Ch3/Ch2 and Ch4/Ch2 around the G8.14+0.23 region,
respectively. The ratio contours are also overlaid on the ratio maps for better clarity and insight (see Fig.~\ref{fig3}). 
Both the ratio maps clearly trace the prominent PAH emissions and subsequently extent of PDRs in the region.
 
\begin{figure}
\plotone{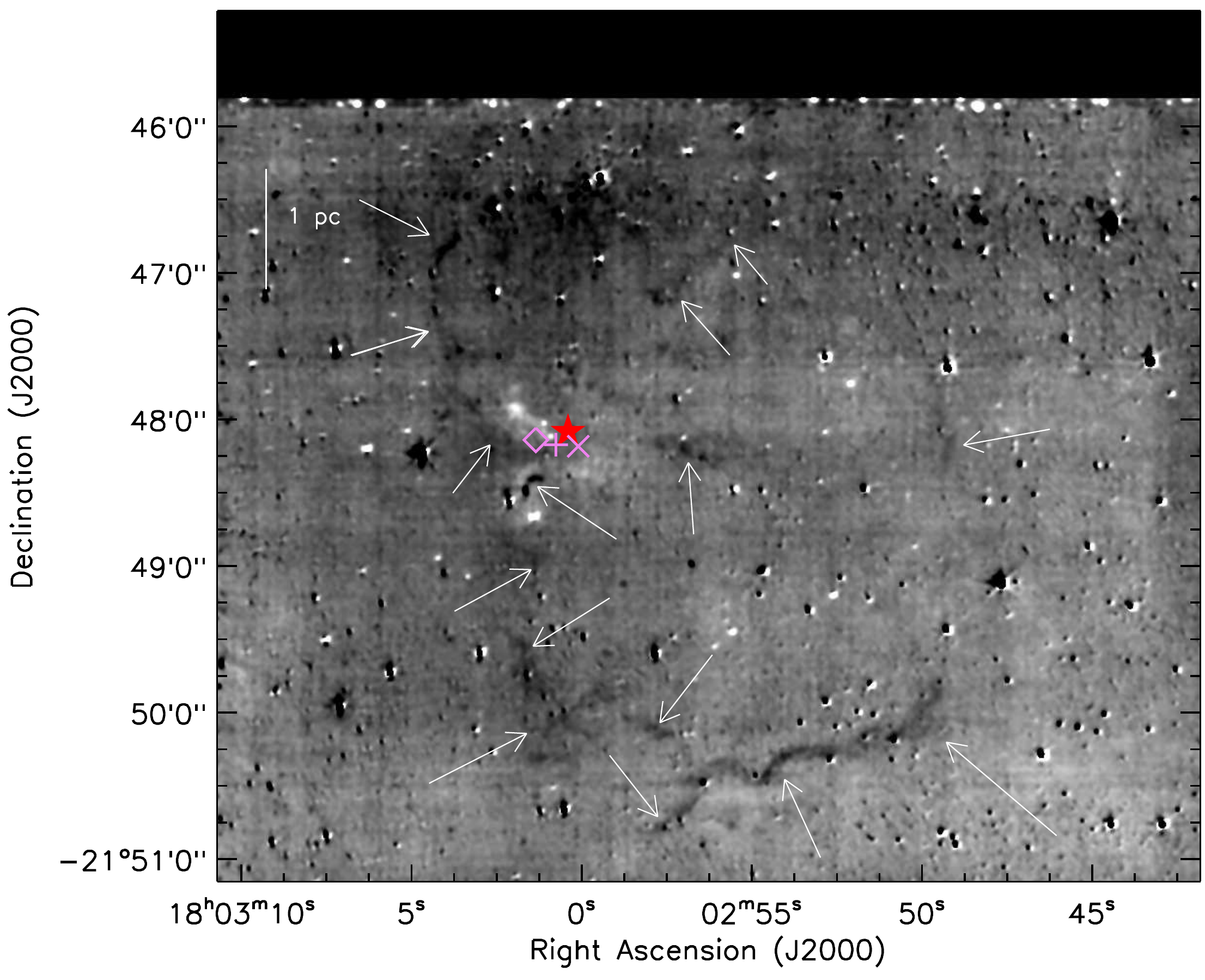}
\caption{Figure represents the inverted grey-scale image of the continuum-subtracted H$_{2}$ image at 2.12 $\mu$m.
White arrows indicate the detected H$_{2}$ emission along the bubbles (as seen in GLIMPSE images) around the G8.14+0.23 region. The other
marked symbols are similar to as shown in Fig.~\ref{fig1}. The continuum-subtracted H$_{2}$ image is processed to median filtering
with a width of 4 pixels and smoothened by 4 $\times$ 4 pixels using boxcar algorithm to trace out the faint features in the image. \label{fig7}}
\end{figure}

Figure~\ref{fig4} shows the IRAC Ch2/Ch1 ratio map overlaid with Ch3/Ch2 ratio and MAGPIS 20 cm radio contours.
The bright region traced by the Ch2/Ch1 ratio map is coincident with the 20 cm radio continuum emission, possibly due to presence of the 
Br$\alpha$ (4.05 $\mu$m) feature in Ch2 band around the H\,{\sc ii} region. One can also notice very faint diffuse emission 
along the edges of the bubbles away from the peak of 20 cm radio emission, possibly due to the molecular hydrogen feature in Ch2 band. 
\citet{ojeda02} detected two sources in the 6 cm radio observations around 
the G8.14+0.23 region. The positions of these sources are also shown in Figure~\ref{fig4}. 
Figures~\ref{fig5}a,~\ref{fig5}b \&~\ref{fig5}c show the grey-scale images in SCUBA 850 $\mu$m, 
BOLOCAM 1.1 mm and JCMT CO 3-2 (345.79599 GHz) around the G8.14+0.23 region respectively. 
Figure~\ref{fig5} exhibits the evidence of collected material along the bubbles traced by dust 
continuum (SCUBA 850 $\mu$m \& BOLOCAM 1.1 mm) and molecular gas (JCMT CO 3-2 at 345.79599 GHz) emission, which is also indicated by 
red arrows. These arrows are marked to show the associated faint features above the 4-sigma noise level for each image. 
We further utilized JCMT public processed CO 3-2 cube data to observationally check the association of 
the molecular material with the expanding H\,{\sc ii} region seen as MIR bubbles. 
We calculated the mean H$_{2}$ number density near the H\,{\sc ii} region using $^{12}$CO zeroth moment map. 
We derived the column density using the formula $N_{\rm H_{2}}$ (cm$^{-2}$) = $X$ $\times$ $W_{\rm CO}$, where 
$X$ = 6 $\times$ 10$^{20}$ cm$^{-2}$ K$^{-1}$ km$^{-1}$ s \citep[see][]{ji12} and $W_{\rm CO}$ = 48.08  K km s$^{-1}$ from our map. 
The mean H$_{2}$ number density is obtained to be 3575.7 cm$^{-3}$ using the relation $N_{\rm H_{2}}$ (cm$^{-2}$)/$L$ (cm), 
where $L$ is the molecular 
core size of about 8.07 $\times$ 10$^{18}$ cm ($\sim$ 2.6 pc) near the H\,{\sc ii} region. Figure~\ref{fig6} exhibits molecular emission around the region between 17 and 25 km s$^{-1}$ 
velocity interval in steps of 1 km s$^{-1}$.  
The bubbles (CN 107 and CN 109) are clearly traced by molecular emission in the velocity range of 19 to 21 km s$^{-1}$ (see Fig.~\ref{fig6}).  
We also estimated velocity dispersion of about 4 km s$^{-1}$  (using second moment CO map) around the region, assuming a smooth rotation curve 
for the Galaxy with typical gas dispersion velocity of about 10 km s$^{-1}$ \citep{stark89,silk97}. 
The consideration of velocity ranges of molecular gas in which bubbles are traced (see Fig.~\ref{fig6}) 
and the velocity dispersion value, imply that the molecular gas in these bubbles are physically associated. 
These velocity ranges of molecular gas are also compatible with the ionized gas velocity ($\sim$ 20.3 km s$^{-1}$) obtained by the H76$\alpha$ recombination 
line profile study around the H\,{\sc ii} region \citep{kim01}, which shows the physical association of the molecular material and the H\,{\sc ii} region. 
The evidence of collected material along the bubbles is further confirmed by the detection of the H$_{2}$ emission (see Figure~\ref{fig7}). 
Figure~\ref{fig7} represents the continuum-subtracted H$_{2}$ image at 2.12 $\mu$m and reveals that the H$_{2}$ emission surrounds 
the H\,{\sc ii} region along bubbles, forming a PDR region, which may be collected due to the shock. 
In brief, the PAH emission, cold dust emission, molecular CO gas and shocked H$_{2}$ emissions are coincident along the bubbles.
 
\begin{figure}
\plotone{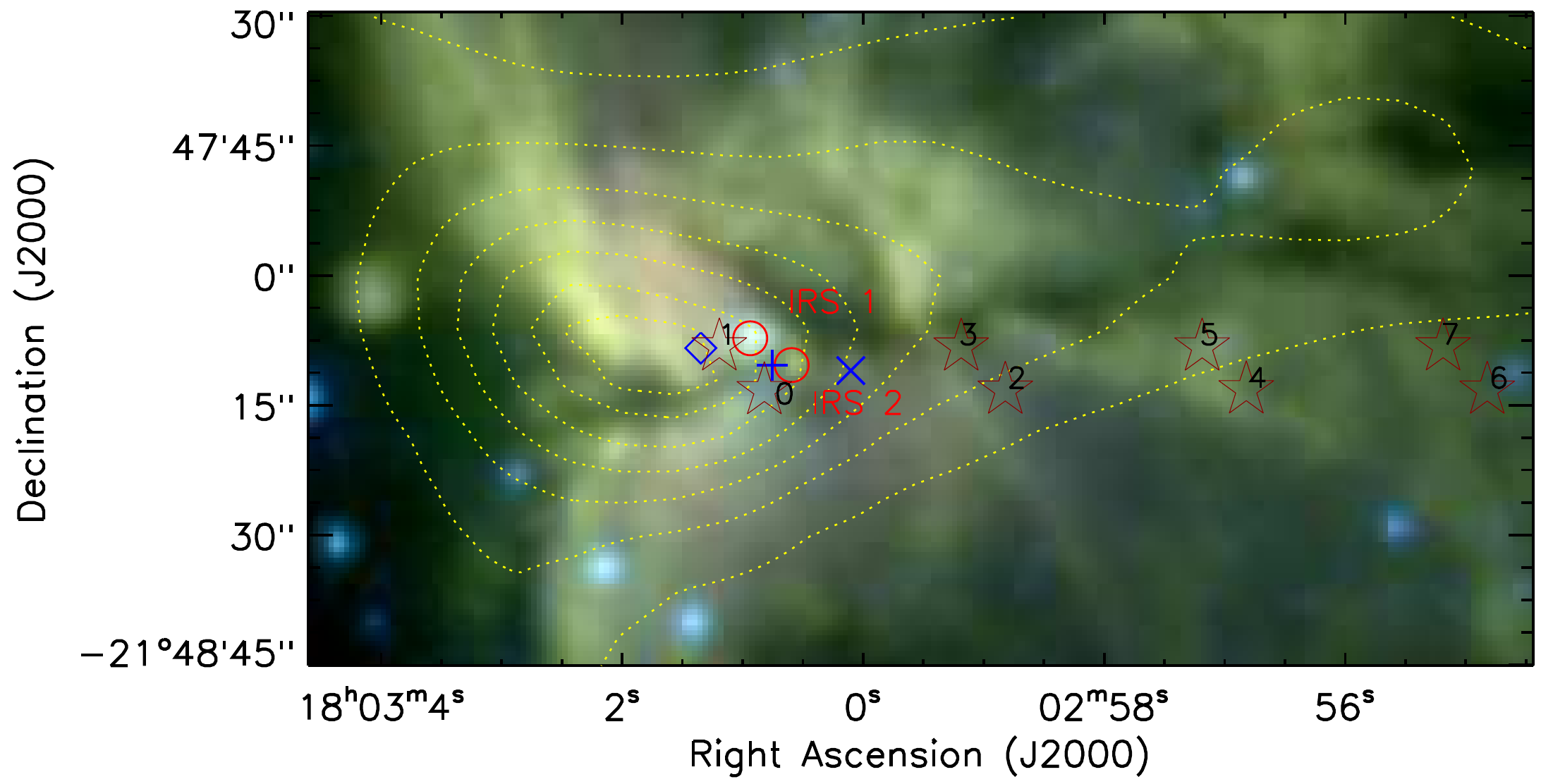}
\caption{Zoomed-in 3 color composite image using IRAC 4.5 $\mu$m (blue), 5.8 $\mu$m (green) and 8.0 $\mu$m (red) around
the G8.14+0.23 region (size $\sim$ 2.36 $\times$ 1.26 arcmin$^{2}$; as shown by a dashed box in Fig.~\ref{fig1}).
Star symbols represent 8 positions of $\it Spitzer$-IRS slit around the G8.14+0.23 region and also labeled as 0 to 7. SCUBA 850 $\mu$m emission is also
shown by dotted yellow contours with similar levels as shown in Fig.~\ref{fig2}.
The positions of IRS~1 \& IRS~2 sources are marked by red circles and labeled in the image. \label{fig8}}
\end{figure}

Figure~\ref{fig8} exhibits the zoomed-in color composite image using IRAC/GLIMPSE images (red: 8.0 $\mu$m, green: 5.8 $\mu$m and blue: 4.5 $\mu$m) 
around the G8.14+0.23 region. The slit positions of $\it Spitzer$-IRS spectrograph at 8 observed locations around the G8.14+0.23 region are also 
marked in the Figure~\ref{fig8} with star symbols \citep[see][]{okada08}. \citet{okada08} also listed the line intensities of detected 
emission lines ([Ar\,{\sc iii}], 22 $\mu$m; [Fe\,{\sc iii}], 23 $\mu$m; [Fe\,{\sc ii}], 26 $\mu$m; H$_{2}$ 0-0 S(0), 28 $\mu$m; 
[S\,{\sc iii}], 33 $\mu$m; [Si\,{\sc ii}], 35 $\mu$m and [Ne\,{\sc iii}], 36 $\mu$m) at the 8 observed positions (similar labels marked 
as 0 to 7 in Figure~\ref{fig8}) around the G8.14+0.23 region. 
The presence of both PAH and fine structure line emissions around the region clearly suggest, the presence of an ionizing 
source with UV radiation close to radio and dust continuum peaks. 

\subsection{Photometric analysis of point-like sources around the G8.14+0.23 region}
In order to trace ongoing star formation activity in the G8.14+0.23 region, we have identified YSOs using NIR and GLIMPSE data.

\subsubsection{Identification of YSOs}
\label{subsec:phot1}
\begin{figure}
\plotone{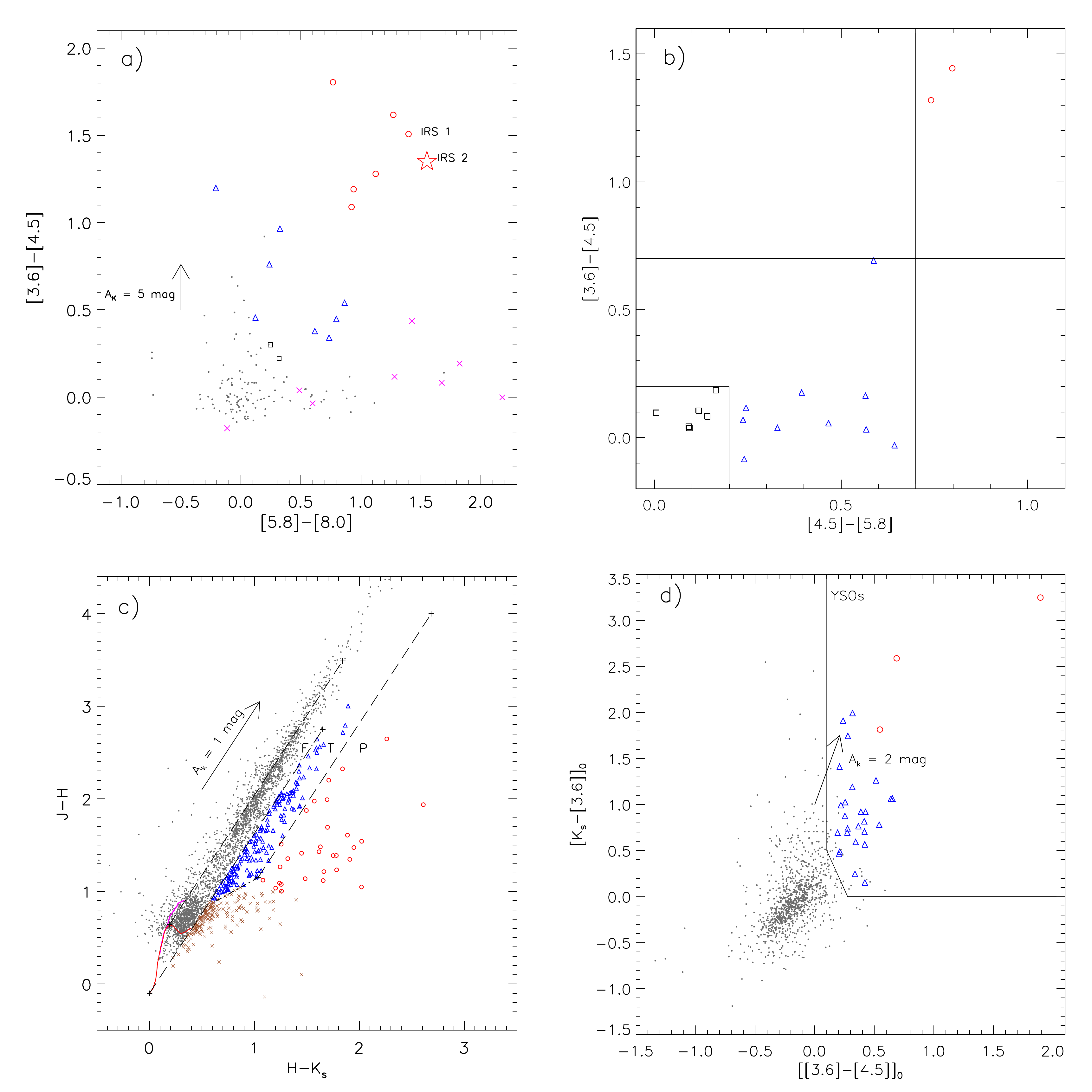}
\caption{\footnotesize\scriptsize a) Color-color diagram (CC-D) using the {\it Spitzer}-IRAC four bands for all the sources identified
within the region shown in Fig.~\ref{fig1}. The extinction vector for A$_{K}$ = 5 mag is shown by an arrow,
using average extinction law from \citet{Flaherty07}. The dots in grey color around the centre (0,0) locate
the stars with only photospheric emissions. The open squares (black), open triangles (blue) and open circles (red)
represent Class III, Class II and Class 0/I sources respectively, classified using the $\alpha_{IRAC}$ criteria.
The ``$\times$'' symbols in magenta color show the identified PAH-emission-contaminated apertures in the region. The source IRS~2 is marked by a red star symbol and is also labeled along with the source IRS~1 in the diagram (see the text).
b) CC-D of the sources detected in three IRAC/GLIMPSE bands, except 8.0 $\mu$m. The open squares (black), open triangles (blue) and
open circles (red) represent Class III, Class II and Class 0/I sources respectively, classified following the criteria
suggested by \citet{hartmann05} and \citet{getman07}. c) UKIDSS JHK$_{s}$ CC-D for the G8.14+0.23 region.
UKIDSS NIR (JHK$_{s}$) magnitudes are calibrated in 2MASS system.
The unreddened dwarf and giants loci are shown by red and magenta curves \citep{bessell88} respectively.
The long-dashed straight lines across the CC-D are the extinction vectors (A$_{K}$ = 3 mag) using \citet{indebetouw05} extinction law.
The extinction vector A$_{K}$ = 1 mag is also shown in the diagram.
Classical T Tauri (CTTS) locus (in California Institute of Technology (CIT) system) \citep{meyer97} is shown by a
dashed-dotted line. The loci of unreddened dwarf (Bessell \& Brett (BB) system), giant (BB-system) and CTTS (CIT system) are converted into 2MASS system using
transformation equations given by \citet{carpenter01}. Red circles and open blue triangles represent the Class I and class II YSOs.
The CC-D is classified into three regions namely ``F'', ``T'', and ``P'' (see text for details).
d) Figure shows the de-reddened [K$_{s}$ - [3.6]]$_{0}$ $vs$ [[3.6] - [4.5]]$_{0}$ CC-D using NIR and GLIMPSE data.
The selected region shown by the solid lines represents the location of YSOs.
The extinction vector for A$_{K}$ = 2 mag is shown by an arrow, calculated using the average
extinction law from \citet{Flaherty07}.
Open red circles and open blue triangles represent Class I and Class II sources respectively. \label{fig9}}
\end{figure}

We have used \citet{gutermuth09} criteria based on four IRAC bands to identify YSOs and various possible 
contaminants (e.g. broad-line AGNs, PAH-emitting galaxies, shocked emission blobs/knots and PAH-emission-contaminated apertures). 
These YSOs are further classified into different evolutionary stages (i.e. Class I, Class II, Class III and photospheres) 
using slopes of the IRAC spectral energy distribution (SED). 
Fig.~\ref{fig9}a shows the IRAC color-color ([3.6]-[4.5] vs [5.8]-[8.0]) diagram for all the identified sources. 
We find 14 YSOs (6 Class 0/I; 8 Class II), 2 Class III, 123 photospheres and 
8 contaminants in the G8.14+0.23 region. Two sources close to the peak of the dense molecular gas and the dust clump 
at the interface of the bubbles, IRS~1 and IRS~2, are also labeled in Fig.~\ref{fig9}a. The position of IRS~2 is shown by a star symbol 
using 3.6 $\mu$m mag as an upper limit (see Table~\ref{tab1}). It is to be noted that IRS~2 is close (about 2 arcsec) to the detected 
methanol maser. The details of the YSO classification can be found in \citet{dewangan11}. 
We have also applied criteria to identify YSOs as suggested by \citet{hartmann05} and \citet{getman07} for those sources that are 
detected in three IRAC/GLIMPSE bands, but not in the 8.0 $\mu$m band. We identify 12 more YSOs (2 Class 0/I; 10 Class II) through 
color-color diagram using three GLIMPSE bands in the region (see Fig.~\ref{fig9}b).

UKIDSS NIR JHK$_{s}$ photometry is used to identify more YSOs in the region. 
NIR color-color diagram (CC-D: H-K$_{s}$/J-H) is shown in Fig.~\ref{fig9}c for all the sources detected in J, H and K$_{s}$ bands. 
We have divided the CC diagram into three regions namely ``F'', ``T'', and ``P'' \citep[e.g.,][]{sugitani02,ojha04a,ojha04b}. 
The ``F'' sources are located between the reddening bands of the main-sequence and giant stars. 
The sources located within the ``T'' region along the T Tauri locus are T Tauri-like sources (Class II objects) with large NIR excess. 
``P'' sources are those located in the region redward of region ``T'', and are most likely Class I objects (protostellar objects). 
We have identified 159 Class~II and 30 Class~I sources located in the ``T'' and ``P'' regions in the Fig.~\ref{fig9}c, 
respectively. The identified Class~I and Class II sources are shown by red circles and blue triangles in Fig.~\ref{fig9}c. 
We have estimated the visual extinction (A$_{V}$) using the CC diagram 
and following the \citet{indebetouw05} extinction law, for those sources that fall within 
the ``F'' region \citep[see][]{ojha10}. The average value comes out to be A$_{V}$ $\sim$ 8.2 mag, which is in 
good agreement with that estimated (A$_{V}$ $\sim$ 8.8 mag) by \citet{rowles09} using 2MASS data for the G8.14+0.23 region.
 
GLIMPSE 3.6 and 4.5 $\mu$m bands are more sensitive for point sources than GLIMPSE 5.8 and 8.0 $\mu$m images.
Therefore, a larger number of YSOs can be identified using a combination of NIR (JHK$_{s}$) with GLIMPSE 3.6 and 
4.5 $\mu$m (i.e. NIR-IRAC) photometry, where sources are not detected in IRAC 5.8 and/or 8.0 $\mu$m band \citep{gutermuth09}. 
We followed the criteria given by \citet{gutermuth09} to identify YSOs using H, K$_{s}$, 3.6 and 4.5 $\mu$m data. 
We have found 29 more YSOs (26 Class II and 3 Class I) using NIR-IRAC data (see Fig.~\ref{fig9}d).
 
Finally, we have obtained a total of 244 YSOs (203 Class II and 41 Class I) using NIR and GLIMPSE data in the region. 

\begin{deluxetable*}{cccccccccccc}
\tablewidth{0pt}
\tabletypesize{\scriptsize}
\setlength{\tabcolsep}{0.05in}
\tablecaption{NIR and {\it Spitzer} IRAC/GLIMPSE photometric magnitudes are listed for selected clustered YSOs
having less than or equal to $d_{c}$ value close to the sub-mm peak (see the text). Spectral index ($\alpha_{IRAC}$) was
obtained by fitting of IRAC wavelengths (sources having atleast 3 IRAC wavelengths magnitude). \label{tab1}}

\tablehead{ \colhead{Source} & \colhead{RA[2000]} & \colhead{DEC[2000]} & \colhead{J (mag)} & \colhead{H (mag)}
& \colhead{K$_{s}$ (mag)} & \colhead{Ch1 (mag)} & \colhead{Ch2 (mag)} &
\colhead{Ch3 (mag)} & \colhead{Ch4 (mag)} &
\colhead{$\alpha_{IRAC}$}  }
\startdata
     IRS~1        &  18:03:00.94    &  -21:48:07.3   &         ---       &         ---       & 13.44$\pm$0.01   &      8.79$\pm$0.01   &      7.29$\pm$0.01  &   6.07$\pm$0.03    &  4.67$\pm$0.04   &   1.85   \\
     IRS~2        &  18:03:00.59    &  -21:48:10.5   &         ---       &         ---       &        ---       &     $\ll$ 11.78      &     10.39$\pm$0.10  &   7.68$\pm$0.05    &  6.13$\pm$0.07   &   3.96   \\
     IRS~3        &  18:03:00.36    &  -21:48:23.5   &  15.97$\pm$0.01   &  13.76$\pm$0.00   & 12.50$\pm$0.00   &     11.44$\pm$0.08   &     10.83$\pm$0.09  &         ---        &        ---       &    --    \\
     IRS~4        &  18:03:01.12    &  -21:48:01.9   &  16.63$\pm$0.02   &  14.98$\pm$0.01   & 13.54$\pm$0.01   &     11.10$\pm$0.14   &      9.43$\pm$0.10  &         ---        &        ---       &    --    \\
     IRS~5        &  18:03:02.90    &  -21:48:23.1   &         ---       &         ---       & 15.27$\pm$0.03   &     11.54$\pm$0.02   &      9.92$\pm$0.02  &   8.71$\pm$0.04    &  7.44$\pm$0.06   &   1.81   \\
     IRS~6        &  18:03:03.70    &  -21:48:45.8   &  17.94$\pm$0.05   &  15.87$\pm$0.02   & 14.51$\pm$0.01   &     13.14$\pm$0.05   &     12.59$\pm$0.07  &        ---         &       ---        &   --    \\
     IRS~7        &  18:03:04.33    &  -21:47:46.1   &  15.20$\pm$0.01   &  14.39$\pm$0.01   & 13.88$\pm$0.01   &     13.44$\pm$0.08   &     13.05$\pm$0.10  &        ---         &       ---        &   --    \\
     IRS~8        &  18:03:04.39    &  -21:48:31.1   &        ---        &        ---        & 15.72$\pm$0.04   &     11.27$\pm$0.01   &      9.47$\pm$0.01  &   8.55$\pm$0.01    &  7.79$\pm$0.02   &  1.03   \\
\enddata
\end{deluxetable*}

\subsubsection{YSOs surface density map}
\label{subsec:surfden}
\begin{figure}
\plotone{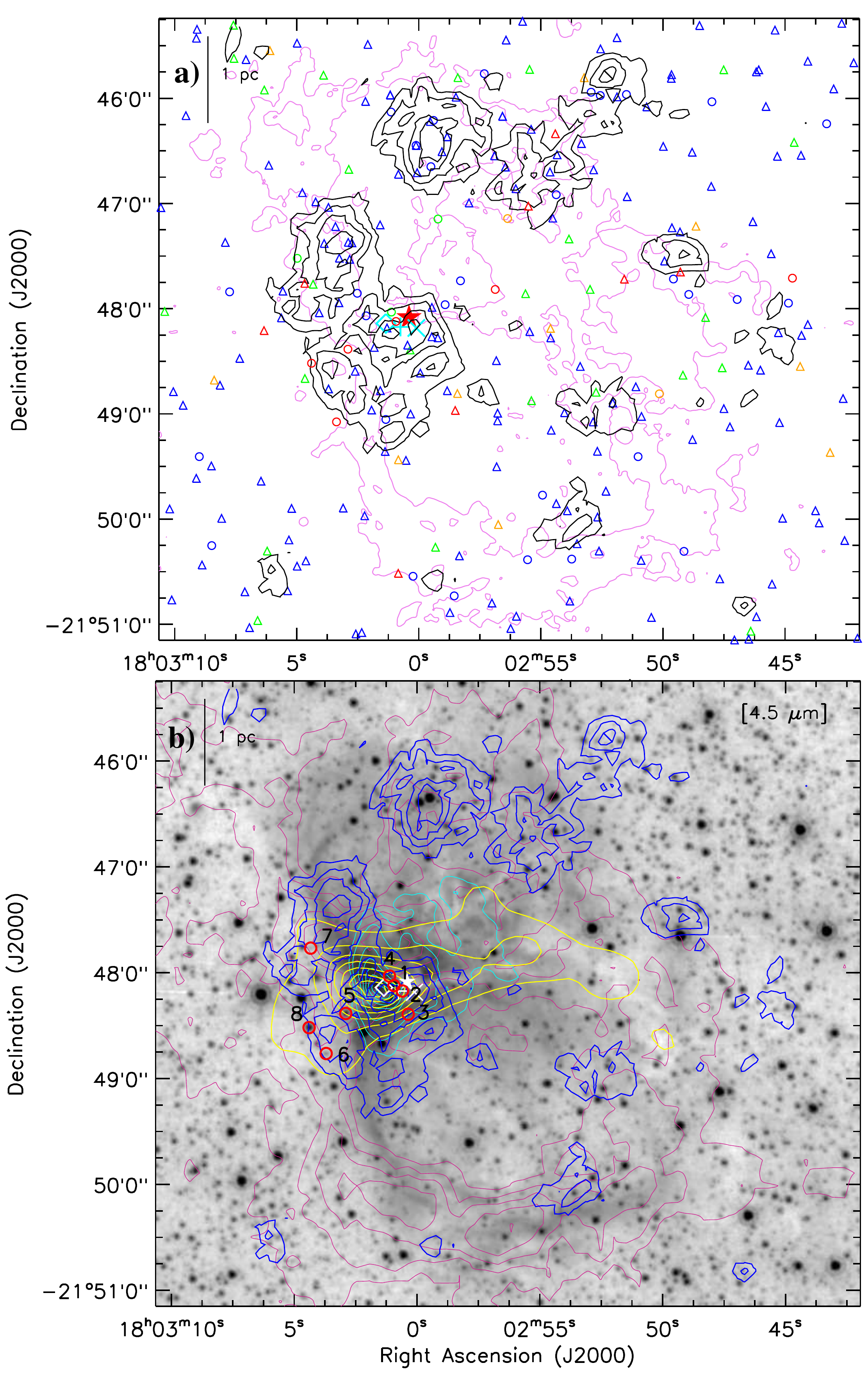}
\caption{\footnotesize\scriptsize a) Figure represents the spatial distribution of all identified YSOs in the G8.14+0.23 region.
The YSO surface density contours are plotted for 6, 8, 11 and 16 YSOs/pc$^{2}$, from outer to inner side (see text for details).
The open circles and open triangles show the Class I and Class II sources respectively.
The YSOs identified using four IRAC, three IRAC, NIR-IRAC and NIR data are shown by red, orange, green and blue colors respectively.
The Ch3/Ch2 ratio contours in violet color are also overlaid on the image with a similar level as in Fig.~\ref{fig4} and
other marked symbols are similar to as shown in Fig.~\ref{fig1}.
b) The schematic represents the spatial distribution of YSOs, dust emission, molecular gas and ionized gas in the G8.14+0.23 region.
GLIMPSE 4.5 $\mu$m image is overlaid with contours of YSO surface density (blue; similar levels as in Fig.~\ref{fig10}a),
SCUBA 850 $\mu$m (yellow), MAGPIS 20 cm (cyan) and JCMT CO 3-2 (violet red) data.
The CO 3-2 contour levels are 20, 29, 40, 55, 70 and 80 \% of the peak i.e. 139.551 K km s$^{-1}$.
The contour levels of SCUBA 850 $\mu$m and MAGPIS 20 cm are similar to as shown in Fig.~\ref{fig2}.
The selected clustered YSOs (see subsection 3.3.3) are also marked (red circles) and labeled as 1,....,8
on the image (see Table ~\ref{tab1}). \label{fig10}}
\end{figure}

To study the spatial distribution of YSOs, we generated the surface density map of all YSOs using a 5 arcsec grid size, 
following the same procedure as given in \citet{gutermuth09}. 
The surface density map of YSOs is constructed using 6 nearest-neighbour (NN) YSOs for each grid point. 
Fig.~\ref{fig10}a shows the spatial distribution of all identified YSOs (Class I and Class II) in the region. 
The contours of YSO surface density and Ch3/Ch2 ratio map are also overlaid on the diagram. 
The levels of YSO surface density contours are 6 (2.5$\sigma$), 8 (3.3$\sigma$), 11 (4.6$\sigma$) and 16 (6.7$\sigma$) YSOs/pc$^{2}$, increasing from 
the outer to the inner region. We have also calculated the empirical cumulative distribution (ECD) as a function of NN distance  
to identify the clustered YSOs in the region. Using the ECD, we estimate the distance of 
inflection $d_{c}$ = 0.617 pc (0.0084 degrees at 4.2 kpc) for the region for a surface density of 
6 YSOs/pc$^{2}$ \citep[see][for details of $d_{c}$ and ECD]{dewangan11}. 
We find that 37\% (91 out of 244 YSOs) YSOs are present in clusters. 
Figure~\ref{fig10}a reveals that the distribution of YSOs is mostly concentrated in the North 
and East regions, having peak density of about 20 YSOs/pc$^{2}$, while YSO density of about 8 (3$\sigma$) YSOs/pc$^{2}$ is 
also seen around the West and South-East regions (see Figure~\ref{fig10}a). 
These values of YSO surface density is much lower than usual values (about 30--60 YSOs/pc$^{2}$) found in recent 
works \citep[see][]{gutermuth09,gutermuth11} in MSF regions. 
Also the percentage of YSOs in clusters is low (only 37\%) as compared to recent reports (about 60--80\%) in 
MSF regions \citep[see][]{gutermuth09,gutermuth11,chavarria08}. 
It possibly supports the youth of the region and a site of further ongoing star formation. 
It is also to be noted that the YSO surface density is associated with the PDR region, on the edge of bubbles. The correlation of 
cold dust, molecular gas, ionized gas and YSO surface density is shown in Figure~\ref{fig10}b. The association of YSOs with the 
collected materials around the region further reveals the ongoing star formation on the edge of bubbles.\\ 
In addition, we have checked the possibility of intrinsically ``red sources'' contamination, such as AGB stars 
in our YSO sample. Recently, \citet{Robitaille08} prepared an extensive catalogue of such red sources based 
on the {\it Spitzer} GLIMPSE and MIPSGAL surveys. They showed that two classes of sources are well separated in the [8.0 - 24.0] color 
space such that YSOs are redder than AGB stars in this space \citep[see also][]{Whitney08}. 
Since, MIPS 24 $\mu$m data is saturated for the G8.14+0.23 region, we have therefore used criteria 
based on the IRAC magnitude (4.5 $\mu$m) and color space ([4.5 - 8.0]), to estimate the 
AGB contamination. It is believed that the presence of AGB stars in the cluster is 
unlikely \citep[see][and references therein]{dewangan11}. Therefore, we have applied red source criteria for 
only those YSOs which are cluster members with less than or equal to $d_{c}$ value (see Fig.~\ref{fig10}a). 
We find that GLIMPSE YSOs may be contaminated by AGB stars up to about 36\% (5 out of 14 YSOs; see subsection~\ref{subsec:phot1}).

\subsubsection{SED modeling of clustered YSOs}
\label{subsec:sed}
\begin{figure}
\plotone{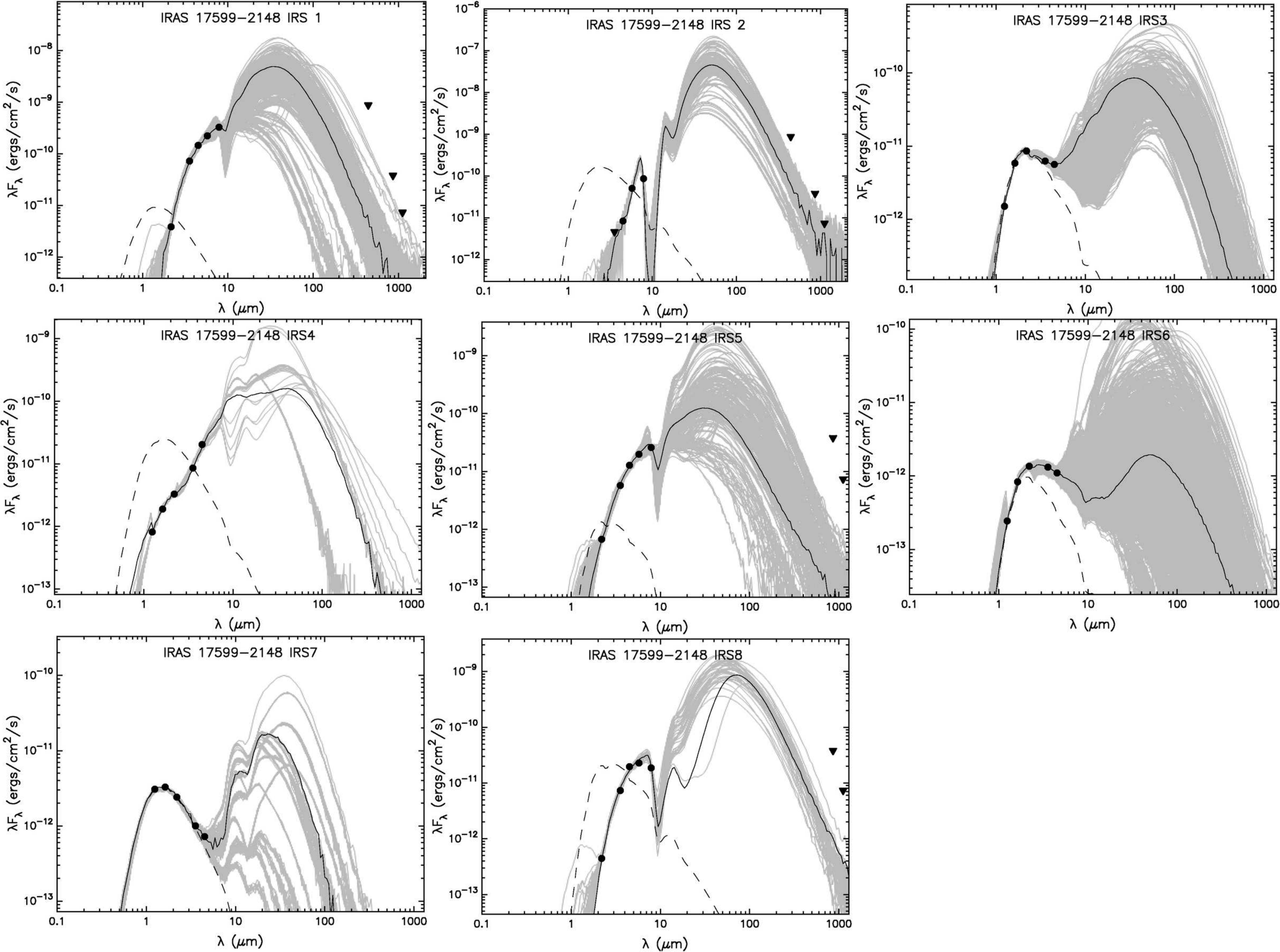}
\caption{The SED plots of all the selected sources including IRS~1 \& IRS~2 are shown here.
Filled circles are observed fluxes of good quality (with filled triangles as upper limits) taken from the archives or published literature (see
text for more details) and the curves show the fitted model with criteria $\chi^{2}$ - $\chi^{2}_{best}$ $<$ 3.
The thin black curve corresponds to the best fitting model. The dashed curves represent
photospheric contributions. \label{fig11}}
\end{figure}

In this subsection, we present SED modeling of some selected YSOs to derive their 
various physical parameters using an on-line SED modeling tool \citep{Robit06,Robit07}. 
Clustering analysis of identified YSOs reveals that there is a grouping of YSOs ($\sim$ 20 YSOs/pc$^{2}$) 
close to the peak of the dense molecular gas and the dust clump in the region (see Fig.~\ref{fig10}b). 
We have selected 8 YSOs (designated as IRS 1,.....,IRS 8) from this cluster close to the dense clump, which are detected atleast 
upto GLIMPSE 4.5 $\mu$m or longer wavelengths for SED modeling. 
NIR and {\it Spitzer} IRAC/GLIMPSE photometric magnitudes for these 
selected YSOs are listed in Table~\ref{tab1} along with IRAC spectral indices ($\alpha_{IRAC}$) and are also labeled in 
Fig.~\ref{fig10}b. IRAC spectral indices were calculated using a least squares fit to the IRAC flux points in a 
log($\lambda$) versus log($\lambda$F$_{\lambda}$) diagram for those sources that are detected in atleast 3 IRAC bands 
\citep[see][for details]{dewangan11}. 
We have not included the MIPS 24 $\mu$m magnitude for SED modeling because the image is saturated 
close to the IRAS position. The SED model tool requires a minimum of three data points with good quality 
as well as the distance to the source and visual extinction value. 
A distance range of 3.7 to 4.7 kpc and the visual extinction in the range from 8 to 25 mag are used as input parameters 
in SED modeling tool for each source to constraint the SEDs. 
These models assume an accretion scenario with a central source associated with 
rotationally flattened infalling envelope, bipolar cavities, and a flared accretion disk, all under radiative 
equilibrium. The model grid consists of 20,000 models of two-dimensional Monte Carlo simulations of radiation 
transfer with 10 inclination angles, resulting in a total of 200,000 SED models. The grid of SED models covers 
the mass range from 0.1 to 50 M$_{\odot}$. Only those models are selected that satisfy the 
criterion $\chi^{2}$ - $\chi^{2}_{best}$ $<$ 3, where $\chi^{2}$ is taken per data point. The plots of SED 
fitted models are shown in Fig.~\ref{fig11} for all the selected sources. 
The weighted mean values of the physical parameters (age, mass, temperature, luminosity, 
envelope accretion rate ($\dot{M}_{env}$) and disk accretion rate ($\dot{M}_{disk}$)) 
along with the standard deviations derived from the SED modeling for all the selected sources are given in Table~\ref{tab2}. 
The table also contains the model-derived weighted mean values of A$_{V}$ with standard deviations and the degeneracy of the 
models (i.e., the number of models that satisfy the $\chi^{2}$ criterion as mentioned above).
The derived SED model parameters show that the average value of mass, age and A$_{V}$ of these clustered 
YSOs are about 7.8 M$_{\odot}$, 0.12 Myr and 15.5 mag, respectively. 
It is interesting to note that the embedded IRS~1 and IRS~2 sources are young and massive protostars. 
Finally, the SED modeling results favour the ongoing star formation around the region with detection of 
young YSOs as well as some massive candidates in their early phase of formation.

\begin{deluxetable*}{cccccccccc}
\tablewidth{0pc}
\tabletypesize{\scriptsize}
\tablecaption{Physical parameters derived from SED modeling of the selected clustered YSOs (see the text). \label{tab2}}

\tablehead{ \colhead{Source} & \colhead{Age} & \colhead{$M_*$} & \colhead{$T_*$} & \colhead{$L_*$}
& \colhead{$\dot{M}_{env}$} & \colhead{$\dot{M}_{disk}$} & \colhead{A$_{V}$} &
\colhead{Degeneracy/}\\
 \colhead{Name} & \colhead{log(yr)} & \colhead{(M$_{\odot}$)} & \colhead{log(T(K))} & \colhead{log(L$_{\odot}$)}
& \colhead{log(M$_{\odot}$ yr$^{-1}$)} & \colhead{log(M$_{\odot}$ yr$^{-1}$)} & \colhead{(mag)} & \colhead{No. of models}}
\startdata
IRS~1    &     5.17$\pm$0.72  &   10.33$\pm$2.11   &   4.23$\pm$0.26  &  3.74$\pm$0.26   &    -4.19$\pm$2.06    &    -6.96$\pm$1.28  &    19.55$\pm$5.47  &      316 \\
IRS~2    &     3.86$\pm$0.24  &   22.40$\pm$2.61   &   4.01$\pm$0.16  &  4.49$\pm$0.20   &    -2.96$\pm$0.12    &    -5.97$\pm$0.51  &    17.63$\pm$2.48  &      158 \\
IRS~3    &     4.92$\pm$0.37  &    3.51$\pm$1.07   &   3.66$\pm$0.03  &  1.88$\pm$0.18   &    -4.46$\pm$0.64    &    -7.70$\pm$1.09  &    13.19$\pm$2.08  &      821 \\
IRS~4    &     5.46$\pm$0.86  &    6.75$\pm$2.53   &   4.11$\pm$0.26  &  3.13$\pm$0.56   &    -4.93$\pm$1.84    &    -8.77$\pm$2.71  &    13.63$\pm$3.85  &       40 \\
IRS~5    &     4.50$\pm$1.09  &    5.01$\pm$2.40   &   3.79$\pm$0.27  &  2.53$\pm$0.45   &    -4.58$\pm$2.05    &    -6.36$\pm$1.38  &    19.49$\pm$5.13  &      376 \\
IRS~6    &     5.73$\pm$0.55  &    2.47$\pm$0.97   &   3.69$\pm$0.09  &  1.24$\pm$0.29   &    -6.15$\pm$1.62    &    -8.49$\pm$1.68  &    14.12$\pm$2.12  &     3271 \\
IRS~7    &     6.43$\pm$0.18  &    4.51$\pm$0.75   &   4.16$\pm$0.09  &  2.52$\pm$0.24   &    -7.90$\pm$0.17    &   -12.63$\pm$0.83  &     8.13$\pm$0.10  &      154 \\
IRS~8    &     4.60$\pm$0.59  &    7.80$\pm$1.11   &   3.80$\pm$0.17  &  3.04$\pm$0.23   &    -3.61$\pm$0.31    &    -6.63$\pm$1.34  &    18.00$\pm$5.82  &       48 \\
\enddata
\end{deluxetable*}

\subsection{Star formation scenario}
\begin{figure}
\epsscale{1.0}
\plotone{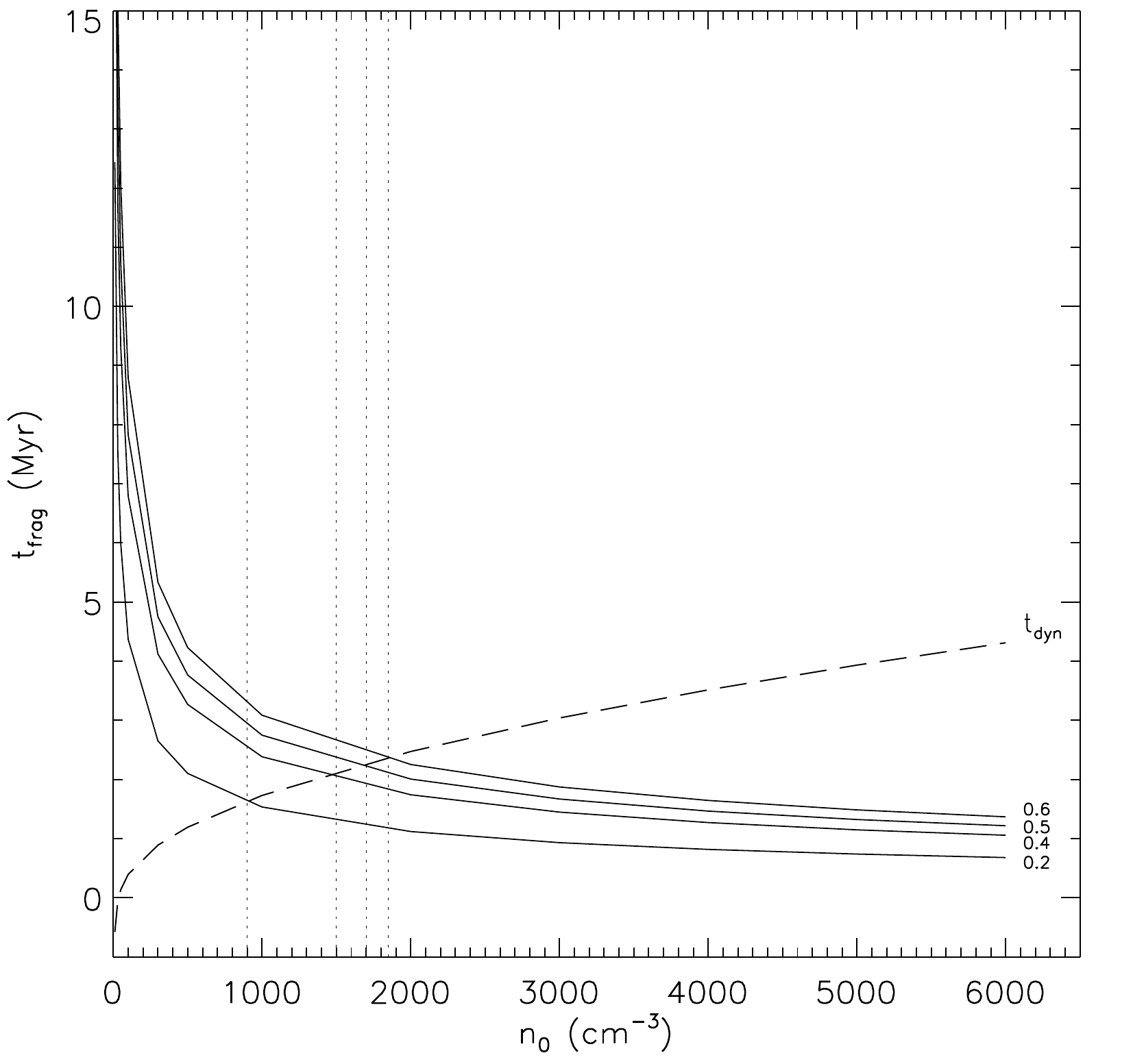}
\caption{The schematic diagram shows the variation of fragmentation time scale (t$_{frag}$) and dynamical time (t$_{dyn}$) with initial density (n$_{0}$) of the
ambient neutral medium. The fragmentation time scale is calculated for different turbulent velocity (a$_{s}$ = 0.2, 0.4, 0.5, \& 0.6 km s$^{-1}$).
Dotted black lines indicate the value of ``n$_{0}$'' at which ``t$_{dyn}$'' is equal to ``t$_{frag}$''
for different ``a$_{s}$'' values. \label{fig12}}
\end{figure}

We have found evidence of collected material along the bubbles and also ongoing formation of YSOs on the edge of the bubbles. 
The morphology and distribution of YSOs suggest that the G8.14+0.23 region is a site of star formation possibly triggered by the expansion of the H\,{\sc ii} region. 
In recent years, the triggered star formation process, especially ``collect and collapse'' mechanism has been studied 
extensively on the edges of many H\,{\sc ii} regions such as Sh 2-104, RCW 79, Sh 2-212, RCW 120, 
Sh 2-217 \citep{deharveng03, deharveng08, deharveng09, zavagno06, zavagno10, brand11}. 
In order to check the ``collect and collapse" process as triggering mechanism around the G8.14+0.23 region, 
we have calculated the dynamical age (t$_{dyn}$) of the H\,{\sc ii} region and compared it with an analytical model by \citet{whitworth94}.
We have estimated the age of the H\,{\sc ii} region at a given radius R, using the following equation \citep{dyson80}:
\begin{equation}
t_{dyn} = \left(\frac{4\,R_{s}}{7\,c_{s}}\right) \,\left[\left(\frac{R}{R_{s}}\right)^{7/4}- 1\right] 
\end{equation}
where c$_{s}$ is the isothermal sound velocity in the ionized gas (c$_{s}$ = 10 km s$^{-1}$) and R$_{s}$ is the radius of the
Str\"{o}mgren sphere, given by R$_{s}$ = (3 N$_{uv}$/4$\pi n^2_{\rm{0}} \alpha_{B}$)$^{1/3}$, where 
the radiative recombination coefficient $\alpha_{B}$ =  2.6 $\times$ 10$^{-13}$ (10$^{4}$ K/T)$^{0.7}$ cm$^{3}$ s$^{-1}$ \citep{kwan97}.
In this calculation, we have used $\alpha_{B}$ = 3.9 $\times$ 10$^{-13}$ cm$^{3}$ s$^{-1}$ for the temperature of 5600 K \citep[see][]{kim01}. 
N$_{uv}$ is the total number of ionizing photons per unit time, emitted by ionizing stars and ``n$_{0}$'' is the initial particle number density of 
the ambient neutral gas. 
We have adopted the Lyman continuum photon flux value (N$_{uv}$ =) of 1.2 $\times$ 10$^{49}$ ph s$^{-1}$ (logN$_{uv}$ = 49.1) from \citet{kim01} 
for an electron temperature, distance and integrated 21 cm (1.43 GHz) flux density of 5600 K, 4.2 kpc, and 6.67 Jy respectively. 
We have estimated the ambient density (n$_{0}$ =) 3575.7 cm$^{-3}$ of the H\,{\sc ii} region 
using $^{12}$CO(J=3-2) line data (see subsection~\ref{subsec:ratmap}). 
Using N$_{uv}$, a mean radius of the H\,{\sc ii} region (R =) 5.9 pc \citep[see][]{kim01}, and n$_{0}$ = 3575.7 cm$^{-3}$, 
we have obtained t$_{dyn}$ $\sim$ 3.3 Myr using Equation~1. Following, \citet{whitworth94} analytical model 
for the ``collect and collapse" process, we have estimated a fragmentation time scale (t$_{frag}$) of 0.86 - 1.73 Myr for 
a turbulent velocity (a$_{s}$ =) of 0.2 - 0.6 km s$^{-1}$ in the collected layer. 
We have found that dynamical age is larger than fragmentation time scale for n$_{0}$ = 3575.7 cm$^{-3}$. 
We have plotted the variation of t$_{frag}$ and t$_{dyn}$ with initial density (n$_{0}$) of the ambient neutral 
medium (see Fig.~\ref{fig12}). The t$_{frag}$ is also calculated for different turbulent velocities in \citet{whitworth94} model. 
It is to be noted that if t$_{dyn}$ is larger than t$_{frag}$, then ambient density (n$_{0}$) should be larger 
than 900, 1500, 1700 and 1850 cm$^{-3}$ for different ``a$_{s}$" values of 0.2, 0.4, 0.5 and 0.6 km s$^{-1}$ respectively (see Fig.~\ref{fig12}). 
We have also estimated the kinematical time scale of molecular bubbles of about 1 Myr ($\sim$ 4 pc/ 4 km s$^{-1}$), 
assuming bubble size of about 4 pc and velocity dispersion $\sim$ 4 km s$^{-1}$ from $^{12}$CO(J=3-2) map (see subsection~\ref{subsec:ratmap}). 
The comparison of the dynamical age of the H\,{\sc ii} region and the kinematical time scale of expanding bubbles with the fragmentation time scale further supports 
triggered star formation and indicates the fragmentation of the molecular materials into clumps due 
to ``collect and collapse'' process around the bubbles. 
Further, the spatial distribution and the clustering analysis of YSOs show the association of YSO clusters 
in the North, East, West and South-East regions with molecular material along the bubbles. 
Among these, the average age of selected 8 YSOs, including IRS~1 and IRS~2 (see subsection~\ref{subsec:sed}),
is about 0.12 Myr (estimated from SED modeling), which is less than the dynamical age of the H\,{\sc ii} region. 
We have also found that these two sources (IRS~1 and IRS~2) are young and massive embedded protostars (about 10 and 22 M$_{\odot}$) 
associated with the dense clump at the interface of the bubbles. It seems that the YSOs present on the edges of the bubbles are 
associated with the collected and fragmented molecular clumps, possibly by the ``collect and collapse'' process \citep[see][]{kang09,pomares09,paron11}. 
This scenario is further supported by the detection of massive YSOs (IRS~1 and IRS~2) associated with the dense clump. 
However, one can not entirely rule out the possibility of triggered star formation by 
compression of the pre-existing dense clumps by the shock wave. It is therefore difficult to distinguish the star formation 
scenario from the pre-existing condensations and/or by the ``collect and collapse'' process in our selected region 
with the available data presented in this work.
 
Recently, \citet{onaka09} reported very broad emission features around 22 $\mu$m in some MSF regions 
(viz., Cas~A, Carina nebula and Sharpless~171) and suggested that this feature possibly originated from the 
dust grains formed in the supernova. Since the G8.14+0.23 region is about 0.3 degrees away from the W30 supernova 
remnant (SNR) and $\it Spitzer$-IRS spectroscopic observations are available for this region in the wavelength 
range of 20 -- 36.5 $\mu$m, it is possible to explore the signature of the 
supernova-induced star formation activity around this region. Following \citet{onaka09}, we therefore re-examined 
the {\it Spitzer}-IRS processed archival spectra (AOR:14928896; PI: Yoko Okada) obtained from ``{\it Spitzer} Heritage 
Archive\footnote[3]{see http://irsa.ipac.caltech.edu/applications/Spitzer/SHA/}'' around the G8.14+0.23 region 
\citep[see IRS slit positions in Fig.~\ref{fig8} and also][]{okada08}. 
We however did not find any broad feature around 22 $\mu$m \citep[see also Table 2 of][]{okada08}. Hence, 
we may reject the possibility of the supernova induced star formation activity in the G8.14+0.23 region. 
\citet{ojeda02} studied 9 MSF regions, including G8.14+0.23, in the vicinity of the 
W30 SNR using 6 cm radio observation and they also ruled out the possibility of a supernova induced star 
formation based on the 6 cm detection around this region and age consideration of the W30 SNR.

\section{Conclusions}
\label{sec:conc}
We have explored the triggered star formation scenario around two bubbles associated with a southern 
massive star forming region G8.14+0.23 using multi-wavelength observations. 
We find that there is a clear evidence of collected material (molecular and cold dust) along the bubbles around the G8.14+0.23 region. 
The JCMT CO velocity map reveals that the molecular gas in the bubbles is physically associated around the region G8.14+0.23. 
The surface density of YSOs reveals ongoing star formation and clustering of YSOs associated with the edge of the bubbles.
We conclude that the YSOs are being formed on the edge of the bubbles possibly by the expansion of the H\,{\sc ii} region. 
We further investigated the ``collect-and-collapse'' process for triggered star formation around the G8.14+0.23 region using 
analytical model of \citet{whitworth94}. We have found that the dynamical age ($\sim$ 3.3 Myr) of 
the H\,{\sc ii} region is larger, and
the kinematical time scale of bubbles ($\sim$ 1 Myr) is comparable to the fragmentation 
time scale ($\sim$ 0.86 - 1.73 Myr) of accumulated gas layers in the region for 3575.7 cm$^{-3}$ ambient density. 
The comparison of the dynamical age and the kinematical time scale of expanding bubbles with the fragmentation time scale 
further supports triggered star formation and indicates the fragmentation of the molecular materials into clumps possibly due 
to the ``collect and collapse'' process around the G8.14+0.23 region. 
We have also investigated infrared counterparts of two young massive embedded protostars 
(about 10 and 22 M$_{\odot}$) associated with dense clumps at the interface of the bubbles. 
It seems that the expansion of the H\,{\sc ii} region is also leading to the formation of these two young massive embedded 
YSOs in the G8.14+0.23 region. Our study possibly favours the ``collect and collapse'' scenario for the formation of YSOs around 
the bubbles in the G8.14+0.23 region. However, we can not entirely rule out the possibility of triggered star formation by 
compression of the pre-existing dense clumps by the shock wave.
  
\acknowledgments
We thank the anonymous referee for a critical reading of the paper and several useful
comments and suggestions, which greatly improved the scientific content of the paper.
This work is based on data obtained as part of the UKIRT Infrared Deep Sky Survey and UWISH2 survey. This publication 
made use of data products from the Two Micron All Sky Survey (a joint project of the University of Massachusetts and 
the Infrared Processing and Analysis Center / California Institute of Technology, funded by NASA and NSF), archival 
data obtained with the {\it Spitzer} Space Telescope (operated by the Jet Propulsion Laboratory, California Institute 
of Technology under a contract with NASA). We thank Dirk Froebrich for providing the narrow band H$_{2}$ image through 
UWISH2 survey.

\end{document}